\documentclass[%
 reprint,
 amsmath,amssymb,
 aps,
superscriptaddress
]{revtex4-2}

\usepackage[dvipdfmx]{graphicx}
\usepackage{dcolumn}
\usepackage{bm}
\usepackage{multirow}
\usepackage{comment}
\usepackage{array,booktabs}
\newcommand{\argmax}{\mathop{\rm arg~max}\limits}

\usepackage{color}
\usepackage{ulem}

\newcommand{\add}[1]{\textcolor{red}{#1}}
\newcommand\erase{\bgroup\markoverwith{\add{\rule[.5ex]{2pt}{0.4pt}}}\ULon}

\begin{document}

\preprint{APS/123-QED}

\title{Lattice protein design using Bayesian learning}

\author{Tomoei Takahashi}
\email{takahashi@phys.cs.i.nagoya-u.ac.jp}
\affiliation{Graduate School of Informatics, Nagoya University, Nagoya, 464-8601, Japan.}
\author{George Chikenji}%
\affiliation{%
Graduate School of Engineering, Nagoya University, Nagoya, 464-8603, Japan.
}%


\author{Kei Tokita}
\affiliation{Graduate School of Informatics, Nagoya University, Nagoya, 464-8601, Japan.}



\date{\today}

\begin{abstract}
Protein design is the inverse approach of the three-dimensional (3D) structure prediction for elucidating the relationship between the 3D structures and amino acid sequences.  In general, the computation of the protein design involves a double loop: a loop for amino acid sequence changes and a loop for an exhaustive conformational search for each amino acid sequence. Herein, we propose a novel statistical mechanical design method using Bayesian learning, which can design lattice proteins without the exhaustive conformational search. We consider a thermodynamic hypothesis of the evolution of proteins and apply it to the prior distribution of amino acid sequences.  Furthermore, we take the water effect into account in view of the grand canonical picture. As a result, on applying the 2D lattice hydrophobic-polar (HP) model, our design method successfully finds an amino acid sequence for which the target conformation has a unique ground state. However, the performance was not as good for the 3D lattice HP models compared to the 2D models. The performance of the 3D model improves on using a 20-letter lattice proteins. Furthermore,  we find a strong linearity between the chemical potential of water and the number of surface residues, thereby revealing the relationship between protein structure and the effect of water molecules. The advantage of our method is that it greatly reduces computation time, because it does not require long calculations for the partition function corresponding to an exhaustive conformational search.  As our method uses a general form of Bayesian learning and statistical mechanics and is not limited to lattice proteins, the results presented here elucidate some heuristics used successfully in previous protein design methods.
\end{abstract}

\maketitle

\section{INTRODUCTION}
Proteins have important roles in living systems. The complex three-dimensional (3D) structure that determines the function of a protein represents an equilibrium state determined by the amino acid sequence and the appropriate physiological conditions\cite{anfinsen1973principles}. Protein design\cite{coluzza2017computational} requires determining the optimal amino acid sequence that results in a given 3D structure as an equilibrium state. Protein design, therefore, is the inverse problem of 3D structure prediction. In addition, as amino acid sequences are determined by genomic information, protein design can be used to explore the design principles of life. In recent decades, many computational protein design methods have been proposed and applied to drug design\cite{dahiyat1997novo, kuhlman2003design, fleishman2011computational, koga2012principles, bale2016accurate, silva2019novo, liu2006rosettadesign, huang2011rosettaremodel}. However, there have been fewer theoretical studies of design principles based on statistical mechanics, and few heuristics have been applied to the design of real proteins.

In order to design an optimal sequence, the most reasonable statistical mechanical procedure involves finding a sequence that minimizes the free energy of a given target conformation. As part of this procedure, one needs to carry out a folding simulation to check that the selected sequence folds into the target conformation with high probability every time a candidate optimal sequence is selected, for example, using the negative design method\cite{jin2003novo}.

However, many other design methods stabilize only a target conformation, although they could in principle lower the energy of other conformations. Nevertheless, they successfully find an optimal sequence that minimizes the energy of the target conformation. It is not clear why such design (inverse problem) methods work without an exhaustive search (forward problem) for all possible compact conformations. One of the main purposes of the present study is to investigate the success of the inverse problem without solving the forward problem, which is a highly non-trivial problem in statistical and biological physics.

In this study, we use the lattice model, which is the simplest coarse-grained protein model. Lattice models have been used to elucidate many problems such as characterizing the free energy landscape of protein folding\cite{cieplak2013energy, shi2016characterzing}, an explanation of the cold denaturation\cite{dijk2016consistent}, the effect of mutation of amino acid sequence for the native structure\cite{holzgrafe2011mutation, shi2014effect}, and the analysis of RNA folding energy landscape\cite{chen2020rna}. The minimal model such as the lattice model is, therefore, still adequate for discussing why the natural protein design methods succeed in designing proteins without the conformational search\cite{dahiyat1997novo, kuhlman2003design, fleishman2011computational, koga2012principles, bale2016accurate, silva2019novo, liu2006rosettadesign, huang2011rosettaremodel}

\section{MODEL AND METHOD}
\subsection{The lattice HP model}
In order to address the problem described above, we take a statistical mechanical approach\cite{cocco2018inverse} based on Bayesian learning, using a coarse-grained protein model called the HP model\cite{lau1989lattice}. We use a lattice model in which every amino acid residue is located on a lattice site, and a protein structure is represented by a self-avoiding walk on a 2D or 3D lattice. Although a real protein has 20 types of amino acid, the HP model includes only two types, hydrophobic (H) and polar (P). Here, we consider $N$ residues ${\boldsymbol \sigma} = \{\sigma_{1}, \sigma_{2}, \ldots ,\sigma_{N} |\forall i,\sigma_{i} = \pm 1\}$ on a lattice position $\bm{r} = \{r_{1}, r_{2}, \ldots ,r_{N} \}$, where $i = 1,2, \ldots, N$ $\sigma_{i} = 1$ indicates that the $i$-th residue is an H residue, and $\sigma_{i} = -1$ indicates that it is a P residue.

We assume the energy of the lattice protein is given by
\begin{eqnarray}
\label{HPenergy}
E(\mbox{\boldmath $r$}; {\boldsymbol \sigma}) = \sum_{i<j}U(\sigma_{i}, \sigma_{j})\Delta(r_{i}-r_{j}),
\end{eqnarray}
where $U(\sigma_{i}, \sigma_{j})$ denotes the interaction potential between the monomers $i$ and $j$. We moreover assume the simplest functional form of $U$: $U(1,1) = \epsilon_{1}, U(1,-1) = U(-1,1) = \epsilon_{2}, U(-1,-1) = \epsilon_{3}$, using two types of interaction set, $(\epsilon_{1}, \epsilon_{2}, \epsilon_{3}) = (-1,0,0)$ and $(-2.3,-1,0)$. The definition of the contact energy $\Delta(r_{i}-r_{j})$ is
\begin{eqnarray}
\label{Delta}
\Delta(r_{i}-r_{j}) = \begin{cases}
						1 & \text{ if $r_{i}$ and $r_{j}$ contact each other,}\\
						0 & \text{otherwise,}
						\end{cases}
\end{eqnarray}
where contact between two residues is defined as the case where $|r_{i} - r_{j}| = 1$ but $|i-j| \neq 1$. In this model, therefore, the energy given by Eq. (\ref{HPenergy}) of denatured conformations is always higher than the energy of compact conformations. Equilibrium statistical mechanics have been successfully applied using the HP model, which is similar to the Ising model. For lattice protein models of comparatively small size, several successful theoretical studies have been reported.

\subsection{Related works}
The first and pioneering study of protein design using statistical mechanics and the HP model was minimization of the energy [Eq. (\ref{HPenergy})] of a target conformation performed by Shakhnovich and Gutin (SG)\cite{shakhnovich1993engineering}. Subsequently, Kurosky and Deutsch proposed a design criterion in which the solution of the design problem is a sequence $\boldsymbol \sigma$ that maximizes the Boltzmann distribution $p(\boldsymbol R |\boldsymbol \sigma)$ as the conditional probability\cite{kurosky1995design, deutsch1996new}, where $\boldsymbol R$ is a set of position vectors of the target conformation. We hereafter call $p(\boldsymbol R | \boldsymbol \sigma)$ the `target probability,' and the design criterion maximizing the target energy probability is denoted the MTP criterion.

For the MTP criterion, a solution $\boldsymbol \sigma_{\rm MTP}$ is given by
\begin{eqnarray}
\label{MTPcriterion}
\boldsymbol \sigma_{\rm MTP} = \argmax_{\boldsymbol \sigma}p(\boldsymbol R|\boldsymbol \sigma)
\end{eqnarray}
and
\begin{eqnarray}
\label{CanonicalDist}
p(\boldsymbol R|\boldsymbol \sigma) = \frac{\exp \left[-\beta E(\boldsymbol R; {\boldsymbol \sigma})\right]}{Z_{\beta}({\boldsymbol \sigma})},\\
\label{PartitionfuncCon}
Z_{\beta}(\boldsymbol \sigma) = \sum_{\bm{r}}\exp \left[-\beta E(\bm{r}; \boldsymbol \sigma)\right],
\end{eqnarray}
where $\beta$ is the inverse temperature of the system,  $\bm{r}$ is any compact conformation of the conformation space into which the sequence $\boldsymbol \sigma$ can fold, and $Z_{\beta}({\boldsymbol \sigma})$ is a partition function of the conformation space. In terms of statistical learning, the MTP criterion is the maximum likelihood estimation for the canonical distribution [Eq. (\ref{CanonicalDist})].
The MTP criterion, however, includes $Z_\beta(\boldsymbol \sigma)$. Hence, to obtain $\boldsymbol \sigma_{\rm MTP}$, one must carry out an exhaustive conformational search every time a candidate $\boldsymbol \sigma_{\rm MTP}$ is found.  For such 
conformational searches, very fast and accurate methods are required. Currently available methods include generalized ensemble Monte Carlo methods\cite{iba1998simulation, chikenji1999multi, ShiraiKikuchi2013}.
 Even these methods, however, cannot provide reasonable results for longer chains (more than 100 residues). Thus, design of large and realistic protein models remains impossible, and successful statistical mechanical protein design using the MTP criterion has been reported only for comparatively small lattice proteins and models\cite{kurosky1995design, deutsch1996new, seno1996optimal, micheletti1998protein, irback1998monte, irback1999design, iba1998design, tokita2000dynamical, rossi2000anovel, betancourt2002protein,  zou2003using, wang2004, jiao2006protein, kleinman2006maximum}.
To overcome the above difficulty, Kurosky and Deutsch\cite{deutsch1996new} carried out high-temperature expansion of the free energy $F_{\beta}(\boldsymbol \sigma) = -(1/\beta)\ln{Z_{\beta}(\boldsymbol \sigma)}$ and minimized $F_{\beta}(\boldsymbol \sigma)$ using simulated annealing for a 2D HP model with $N = 10$--$18$. A design method using simulated annealing for both sequence and conformation space was proposed by Seno {\it et al.}\cite{seno1996optimal}. using a 2D lattice HP model with $N=12$ and $16$. The multi-sequence Monte Carlo method proposed by Irb\"{a}ck {\it et al.}\cite{irback1998monte, irback1999design} is an efficient procedure that obtains an optimal sequence excluding bad sequences with low target probability using fluctuation of sequences; this method was used to design a comparatively large 2D HP model ($N = 32$ and $50$) and 3D off-lattice HP model ($N = 20$). The `design equation' method of Iba {\it et al.} was the first application of Boltzmann machine learning to protein design and was used to obtain a correct sequence for several 3D $3\times3\times3$ cubic conformations\cite{iba1998design, tokita2000dynamical}. There are some other design methods with a similar policy to the design equation method: minimization of the relative entropy\cite{wang2004, jiao2006protein} and gradient descent\cite{kleinman2006maximum} to optimize the sequence.

Some statistical mechanical methods of protein design do not use MTP criterion. Coluzza {\it et al.} proposed a method using energy minimization while maintaining a low variance of the 20 types of amino acid residues for some lattice target conformations\cite{coluzza2003designing}. Coluzza also proposed a design method using minimization of both energy and free energy and designed his original off-lattice protein model\cite{coluzza2011coarse}. Other methods that introduced an explicit solvent to the minimization energy, have also been studied\cite{salvi2002design, abeln2008disordered, ni2013interplay, abeln2014simple, bicanco2017role, bianco2019protein, bianco2020in}. These methods are effective, but the reason for their effectiveness cannot be understood in terms of the MTP criterion. We, therefore, propose a hypothesis as a prior distribution of the Bayesian protein generative model, for this problem in the following subsection.

In recent years, many successful applications of deep learning have been reported in various engineering and scientific fields, including protein folding and drug design\cite{AlphaFold, li2014direct, wang2018computational, greener2018design, merk2018novo, gupta2018generative, o2018spin2, button2019automated}. Deep learning for protein design involves learning the relations between a protein conformation and an amino acid sequence using big data from the Protein Data Bank\cite{bernstein1977protein}. Although deep neural networks have been used to successfully predict an optimal sequence for a target conformation\cite{wang2018computational, button2019automated}, we still do not theoretically understand the design principles of machine learning in this context. In the present study, therefore, we apply Bayesian learning to protein design to explore these design principles.

\subsection{Design method by Bayesian learning}
In Bayesian learning, one assumes that data $D$ are generated by conditional probability $p(D|\theta)$ under the parameter $\theta$. By Bayes' theorem, the posterior distribution $p(\theta|D)$ is given by
\begin{eqnarray}
\label{BayesTheorem1}
p(\theta|D) = \frac{p(D|\theta)p(\theta)}{\int p(D|\theta)p(\theta)d\theta},
\end{eqnarray}
where $p(\theta)$ is the prior distribution of $\theta$ if $\theta$ is a continuous random variable. One estimates $\theta$ and predicts unobserved data from the posterior given by Eq. (\ref{BayesTheorem1}). The basic procedure of Bayesian learning is as follows: starting from a highly arbitrary prior $p(\theta)$, the posterior $p(\theta|D)$ is repeatedly updated with the highly objective likelihood function $p(D|\theta)$ using Eq. (\ref{BayesTheorem1}); thus we can finally obtain a precise value of $\theta$.

Here we apply the above procedure to protein design. Let the appearance probability of the given target conformation $\boldsymbol R$ for data $D$ be the target probability $p(\boldsymbol R|\boldsymbol \sigma)$, and let the prior of the HP sequence $\boldsymbol \sigma$ be the parameter $\theta$, $p(\boldsymbol \sigma)$. Then, the posterior of the sequence $p(\boldsymbol \sigma | \boldsymbol R)$ is given by  
\begin{eqnarray}
\label{BayesTheorem}
p({\boldsymbol \sigma}|\boldsymbol R) = \frac{p(\boldsymbol R|\boldsymbol \sigma)p(\boldsymbol \sigma)}{\sum_{\boldsymbol \sigma}p(\boldsymbol R|\boldsymbol \sigma)p(\boldsymbol \sigma)},
\end{eqnarray}
where $\sum_{\boldsymbol \sigma}$ denotes summation over all sequences.

Furthermore, as with previous studies considering the effect of water\cite{ni2013interplay, abeln2014simple, bicanco2017role, bianco2019protein, bianco2020in}, we consider the interactions between two amino acids [Eq. (1)] as well as the interactions between amino acids and water molecules for the energy of a lattice protein. Nevertheless, we do not simply add the solvation term to the energy of a lattice protein [Eq. (1)]; in fact,  we suppose the grand canonical situation. Thus, we propose a design method controlling the number of water molecules that combine with a protein by adjusting its chemical potential.

We therefore consider the following target probability as a grand canonical distribution:
\begin{eqnarray}
\label{GrandCanonicalDist}
p(\boldsymbol R, N_{\rm w}|\boldsymbol \sigma) = \frac{\exp \left[-\beta \left(E(\boldsymbol R; \boldsymbol \sigma)-\mu N_{\rm w}\right) \right]}{\Xi_{\beta, \mu}(\boldsymbol \sigma)},\\
\label{GrandPartitionfuncCon}
\Xi_{\beta, \mu}(\boldsymbol \sigma) = \sum_{N_{\rm w}=0}^{\infty}\sum_{\bm{r}}\exp \left[-\beta \left(E(\bm{r}; \boldsymbol \sigma)-\mu N_{\rm w} \right) \right].
\end{eqnarray}
The definition of the energy $E(\boldsymbol R; \boldsymbol \sigma)$ of the target $\boldsymbol R$ for a given sequence $\boldsymbol \sigma$ is given by Eq. (\ref{HPenergy}), and $\mu$ is the chemical potential of  water.  We assume that one water molecule combines with one P residue, hence $N_{\rm w} = N_{\rm P}(\boldsymbol \sigma)$, where $N_{\rm P}(\boldsymbol \sigma)$ is the number of all P residues.  Therefore, Eqs. (\ref{GrandCanonicalDist}) and (\ref{GrandPartitionfuncCon}) denote the canonical distribution of the Hamiltonian $E(\boldsymbol R;\boldsymbol \sigma)-\mu N_{\rm P}(\boldsymbol \sigma)$ with external field $\mu$. Consequently, we rewrite Eqs. (\ref{GrandCanonicalDist}) and (\ref{GrandPartitionfuncCon}) as follows:
\begin{eqnarray}
\label{GrandCanonicalDistRev}
p(\boldsymbol R| \boldsymbol \sigma) = \frac{\exp \left[-\beta \left(E(\boldsymbol R; \boldsymbol \sigma)-\mu N_{\rm P}(\boldsymbol \sigma)\right) \right]}{\Xi_{\beta, \mu}(\boldsymbol \sigma)}, \\
\label{GrandPartitionfuncConRev}
\Xi_{\beta, \mu}(\boldsymbol \sigma) = \sum_{\bm{r}}\exp \left[-\beta \left(E(\bm{r}; \boldsymbol \sigma)-\mu N_{\rm P}(\boldsymbol \sigma)\right) \right],
\end{eqnarray}
where $N_{\rm P}(\boldsymbol \sigma)$ is obtained by the conditions $N = N_{\rm H} + N_{\rm P}$ and $\sum_{i}\sigma_{i} = N_{\rm H} - N_{\rm P}$ as
\begin{eqnarray}
\label{NumOfP}
N_{\rm P} = \frac{1}{2}(N - \sum_{i}\sigma_{i}).
\end{eqnarray}
 In order to obtain an optimal sequence $\boldsymbol \sigma$ that maximizes $p(\boldsymbol R|\boldsymbol \sigma)$, we have to repeat the exhaustive conformational search in Eq. (\ref{GrandPartitionfuncConRev}) for each trial sequence $\boldsymbol \sigma$. In general, this calculation takes an enormous amount of time.
 
One of the new ideas of the present study is that the prior distribution $p(\boldsymbol \sigma)$ is given by
\begin{eqnarray}
	\label{PriorDist}
p(\boldsymbol \sigma) = \frac{\Xi_{\beta, \mu}(\boldsymbol \sigma)}{\Xi_{\beta, \mu}},
\end{eqnarray}
\vspace{-6mm}
\begin{eqnarray}
\label{GrandPartitionfunc}
\Xi_{\beta, \mu} = \sum_{\boldsymbol \sigma}\sum_{\bm{r}}\exp \left[-\beta \left(E(\bm{r}; \boldsymbol \sigma)-\mu N_{{\rm P}}(\boldsymbol \sigma)\right) \right],
\end{eqnarray}
where $\Xi_{\beta,\mu}$ is the partition function of both conformation and sequence space. The expression of the prior Eq. (\ref{PriorDist}) is based on the following hypothesis: as a result of evolution, the probability $p(\boldsymbol \sigma)$ is proportional to its partition function $\Xi_{\beta,\mu}(\boldsymbol \sigma)$ so that the free energy $F_{\beta, \mu}(\boldsymbol \sigma):= - (1/\beta)\ln \Xi_{\beta, \mu}(\boldsymbol \sigma)$ takes a minimum. Note that Kurosky and Deutsch\cite{kurosky1995design} assumed the equal {\it a priori} weights $p(\boldsymbol \sigma) = 1/N_{\rm s}$, where $N_{\rm s}$ is the number of all HP sequences. By contrast, our method considers the above postulation for the weight of the appearance of sequences, which is reasonable from the viewpoint of thermodynamics and protein evolution.  We obtain posterior distribution $p(\boldsymbol \sigma | \boldsymbol R)$ by substituting Eqs. (\ref{PriorDist}) and (\ref{GrandCanonicalDistRev}) into Eq. (\ref{BayesTheorem}) and by canceling $\Xi_{\beta,\mu}(\boldsymbol \sigma)$ out as follows:
\begin{eqnarray}
\label{PosteriorDist}
p(\boldsymbol \sigma|\boldsymbol R) = \frac{\exp \left[-\beta \left(E(\boldsymbol R; \boldsymbol \sigma)-\mu N_{{\rm P}}(\boldsymbol \sigma)\right) \right]}{\Xi_{\beta,\mu}(\boldsymbol R)},\\
\label{GrandPartitionfuncSeq}
\Xi_{\beta, \mu}(\boldsymbol R) = \sum_{\boldsymbol \sigma}\exp \left[-\beta \left(E(\boldsymbol R; \boldsymbol \sigma)-\mu N_{{\rm P}}(\boldsymbol \sigma)\right) \right],
\end{eqnarray}
where $\Xi_{\beta,\mu}(\boldsymbol R)$ is a partition function of the sequence space corresponding to the given target conformation $\boldsymbol R$. An important point regarding Eq. (\ref{PosteriorDist}) is that it no longer includes $\Xi_{\beta,\mu}(\boldsymbol \sigma)$. Thus, we can efficiently obtain an optimal sequence using Eq. (\ref{PosteriorDist}) simply by summation over $\boldsymbol \sigma$. This design method without $\Xi_{\beta,\mu}(\boldsymbol \sigma)$ is essentially same as a procedure used previously to obtain more realistic protein designs\cite{dahiyat1997novo, kuhlman2003design, fleishman2011computational, koga2012principles, bale2016accurate, silva2019novo, liu2006rosettadesign, huang2011rosettaremodel}.

The final step is to obtain an optimal HP sequence using Eq. (\ref{PosteriorDist}). Nevertheless, the exact calculation of Eq. (\ref{GrandPartitionfuncSeq}) is difficult if the number of residues $N$ is large.  We thus utilize one of the simplest Markov-chain Monte Carlo (MCMC) methods, Gibbs sampling. In this method, the sampling probability of $\sigma_{i}$ of each Monte Carlo step (MCS) is a conditional probability of $\sigma_{i}$ given other random variables. We thus obtain the following sampling probability by substituting Eqs. (\ref{HPenergy}) and (\ref{NumOfP}) into Eq. (\ref{PosteriorDist}). Accordingly, the sampling probability of an H residue ($\sigma_{i} = 1$) or P residue ($\sigma_{i} = -1$) is given by
\begin{eqnarray}
\label{GSProb}
p(\sigma_{i}=\pm1|\boldsymbol R ; \boldsymbol \sigma_{\backslash i})
= \frac{1}{1+\mathrm{e}^{\pm{\beta}\{\Delta E_{i}(\boldsymbol R; \boldsymbol \sigma) + \mu\}}},
\end{eqnarray}
where $\boldsymbol \sigma_{\backslash i} := \{\sigma_{1},\dots,\sigma_{i-1},\sigma_{i+1},\ldots,\sigma_{N}\}$, a vector of all random variables of residues except for the $i$-th residue $\sigma_{i}$, and the double signs correspond. Let $\Delta E_{i}(\boldsymbol R; \boldsymbol \sigma):=\sum_{j\in n(i)}\left(U(1, \sigma_{j}) - U(-1, \sigma_{j})\right)$, where $n(i)$ denotes the set of sites $j$ that are the nearest neighbors of $i$-th site except for those along the chain ($j\neq i-1, i+1$). The random variables $\boldsymbol \sigma_{\backslash i}$ have fixed realizations in the denominator and the numerator of the right-hand side of Eq. (\ref{PosteriorDist}) at every MCS. Thus, the random variables that interact with the $i$-th residue $\sigma_{i}$ remain only on the right-hand side of Eq. (\ref{GSProb}), because those fixed realizations, except for the residues that interact with $\sigma_{i}$, are canceled out in Eq. (\ref{PosteriorDist}). We decide whether each residue is H or P using the expectation $\left<\sigma_{i}\right>$; that is, $\sigma_{i}$ is H if $\left<\sigma_{i}\right>>0$ and P otherwise. We also take the number of MCSs until the estimated value does not change and let the burn-in be the leading $1/5$ of all MCSs. In this study, the inverse temperature is set to $\beta = 10$ for all conformations of all lattice models. On the other hand, we heuristically set the chemical potential $\mu$ in order to design a unique ground state by repeating the design experiment many times. The necessary and sufficient condition for successful design is that the energy given by Eq. (\ref{HPenergy}) of the target conformation and the sequence designed corresponds to a unique ground state of all possible compact conformations.

The formulation of our design method as described so far can be derived by assuming a joint distribution given by 
\begin{eqnarray}
\label{JointDist}
p(\bm{r}; \boldsymbol \sigma) = \frac{\exp \left[-\beta \left(E(\bm{r}; \boldsymbol \sigma)-\mu N_{{\rm P}}(\boldsymbol \sigma)\right) \right]}{\Xi_{\beta,\mu}},
\end{eqnarray}
where $\Xi_{\beta,\mu}$ is given by Eq. (\ref{GrandPartitionfunc}). One can derive the prior [Eq. (\ref{PriorDist})] and the likelihood function (the target probability) given by Eq. (\ref{GrandCanonicalDistRev}) by a marginalization $p(\boldsymbol \sigma) = \sum_{\bm{r}} p(\bm{r}; \boldsymbol \sigma)$ and a relation $p(\boldsymbol R|\boldsymbol \sigma) = \left(p(\bm{r}; \boldsymbol \sigma)/p(\boldsymbol \sigma)\right)|_{\bm{r} = \boldsymbol R}$, respectively. Thus, the hypothesis [Eq. (\ref{PriorDist})] is included in the joint distribution [Eq. (\ref{JointDist})].

\section{RESULT}
\subsection{Enumerable conformations}
First, we tested our design method with comparatively small lattice protein models, for which all compact conformations were enumerable. We designed 2D $N = 3\times3$, $3\times4$, $4\times4$, $5\times5$, and $6\times6$ lattice models, and 3D $N = 2\times2\times3$ and $3\times3\times3$ lattice models.

Native conformations are not necessarily maximally compact. This is because proteins can have low energy if the hydrophobic core is compact enough\cite{YueandDillPNAS}. Therefore, we designed non-maximally compact conformations of 2D $N = 9$, 12, and 16 used in the study by Irb\"{a}ck and Troein\cite{irback2002} to compare the statistical property between maximally compact and not maximally compact conformations. Following \cite{irback2002}, we do not design approximately unfolded conformations without a core. The examples of both $N = 16$ maximally compact and non-maximally compact conformations are shown in Fig. 1.  

The numbers of all conformations $N_{\rm c}$ including those that are not maximally compact conformations and the numbers of all HP sequences $N_{\rm s}$ of these lattice models are shown in Table \ref{tab:Lattice models used and corresponding number of conformations and H/P sequences}. 

The number of maximally compact conformations is the number of all compact self-avoiding walks from which all kind of rotational, reflection, and head-tail symmetrical conformations have been eliminated. The total number of conformations of these lattice models is enumerable; hence, one can confirm whether or not the designed sequence folds into the target conformation as a unique ground state.

Note that not every conformation always has a solution to the design problem. 

\begin{figure}
	\begin{center}
		\includegraphics[width = 8.0cm]{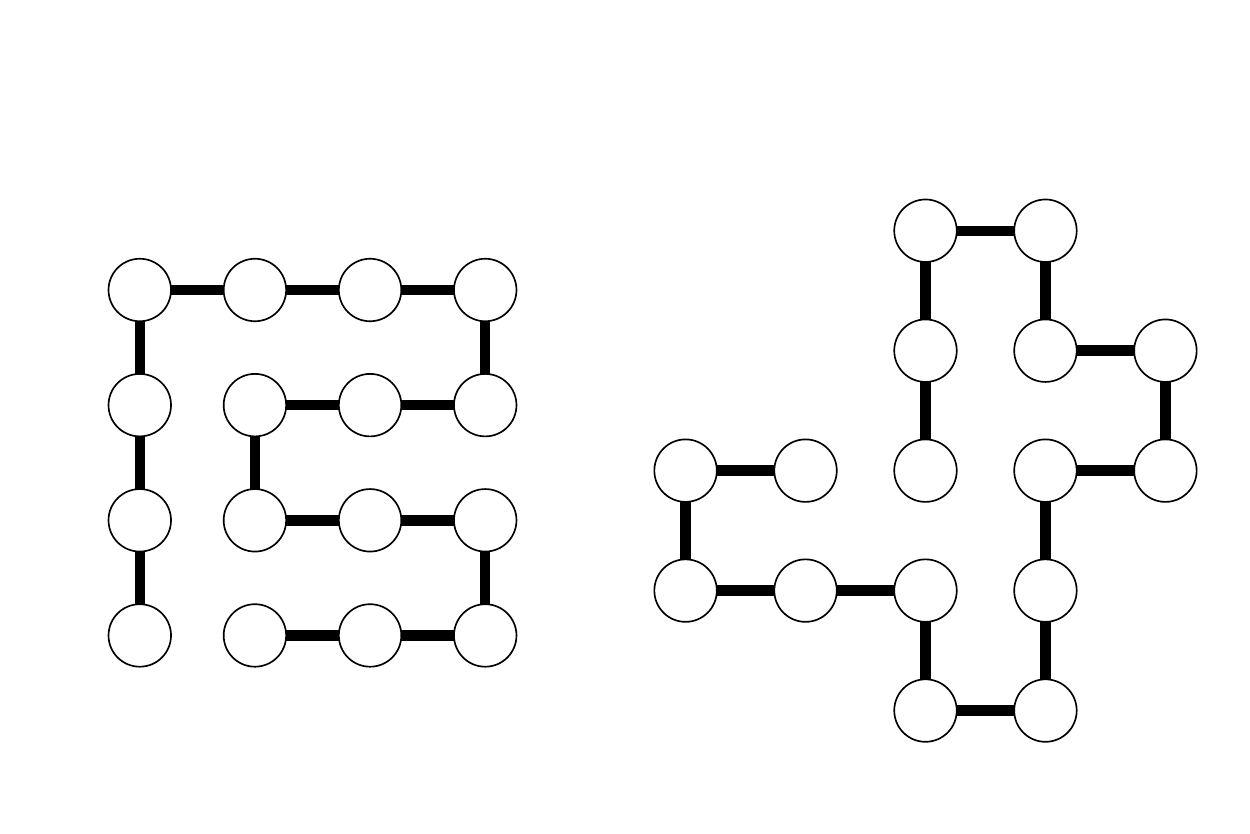}
		\caption{Examples of $N = 16$ maximally compact conformation (left) and non-maximally compact conformation (right). The maximally compact conformation has nine contacts, whereas the non-maximally compact conformation has seven contacts.}
		\label{fig:maximally compact and not maximally compact conformation}
	\end{center}
\end{figure}

The number of sequences that fold into the target conformation as a unique ground state differs among target conformations and is called designability. In general, a conformation with higher designability is easier to design. This is because high designability means a large solution space in sequence space.

Designability is a significant quantity that relates to the thermodynamic stability of proteins; however, we do not address issues of designability in depth here. In order to calculate the exact success rate ($SR$) of the overall conformation, one needs to select designable target conformations with designability greater than zero; however, to enumerate the designabilities of each conformation, one would need to enumerate the energy of every combination of conformations and sequences. This would require vast computation time for models with comparatively large size, such as the $5\times5$, $6\times6$, and $3\times3\times3$ models (Table \ref{tab:Lattice models used and corresponding number of conformations and H/P sequences}), even though they are compact. Therefore, in this study, we carried out the enumeration of designabilities only for the $N = 9$, $N = 12$, $N = 16$, and $2\times2\times3$ lattice models.

For the models with $N=5\times 5$, $6\times 6$, and $3\times 3\times 3$,  the number of conformations was too large. Thus, we randomly chose 100 target conformations and determined the $SR$, that is, the number of successfully designed conformations (Table \ref{tab:Compact lattice models and design results}). For the models with $N=6\times 6$ and $3\times 3\times 3$, we moreover identified the most highly designable conformation (MHDC) (Figs. 1, 2, and 3), in which designabilities were exactly enumerated\cite{li1996emergence}, to test whether our method could be used to design the easiest instance.

\begin{table}[htb]
  \begin{center}
  \caption{Number of conformations and HP sequences. $N = 9$, $N = 12$, and $N = 16$ involves compact $3\times3$, $3\times4$, and $4\times4$ as their maximally compact conformations, respectively. The numbers in brackets represent the numbes of maximally compact conformations, respectively.}
	\begin{tabular}{cccc} \hline \hline
	\addlinespace[2pt]
	Size & $N_{\rm c}$ &$N_{\rm s}$ &Conformations designed\\[3pt] \hline
	\hspace{0.5mm}$N = 9$ & 12 (8) & 512&All\\
	\hspace{0.5mm}$N = 12$ & 52 (27) & 4096&All\\
	\hspace{0.5mm}$N = 16$ & 518 (62) & 65536&All\\
	\hspace{0.5mm}$5\times5$ & 1075 & 33554432&Random 100\\ 
	\hspace{0.5mm}$6\times6$ & 52667 & 68719476736&Random 100 and MHDC\\
	$2\times2\times3$ & 69 & 4096& All\\
	$3\times3\times3$ & 103346 &  134217728& Random 100 and MHDC\\ \hline \hline
	\end{tabular}
	\label{tab:Lattice models used and corresponding number of conformations and H/P sequences}
  \end{center}
\end{table}

The results of the application of our method are summarized in Table \ref{tab:Compact lattice models and design results}. All designed sequences were classified into three types: good, medium, and bad sequences. The good sequences had the target conformation as a unique ground state, medium sequences had the target conformation as one of the degenerated ground states, and bad sequences had ground state conformation(s) that did not include the target conformation. In the table, $SR$, $N_{\rm c}^{(\rm g)}$, $N_{\rm c}^{(\rm m)}$, and $N_{\rm c}^{(\rm b)}$, denote the percentage of good sequences and the number of conformations that were designed with good, medium, and bad sequences, respectively. We also calculated the average degeneracy, $d_{\rm av}$, for all $N_{\rm c}^{(\rm g)} + N_{\rm c}^{(\rm m)}$ ground states. We repeated the calculations with various values of $\mu$ and obtained the optimal value $\mu^*$ that gave the maximum success rate. The values of $\mu^*$ for each lattice size are listed in Table \ref{tab:Compact lattice models and design results}. The values of the energy parameters are also listed in Table \ref{tab:Compact lattice models and design results}. The energy parameters $(\epsilon_{1}, \epsilon_{2}, \epsilon_{3}) = (-2.3, -1,\, 0)$ were also used for a $3\times3\times3$ lattice model in previous work\cite{li1996emergence} in order to avoid the degeneracy of ground states. We used the same energy parameters for 3D lattice models for the same reason. The total MCSs were set to $10^5$ for all target conformations listed in Table \ref{tab:Compact lattice models and design results}. The $N = 9$ and $2\times2\times3$ lattices included several non-designable conformations; we excluded such conformations when calculating $SR$.

According to the results shown in Table \ref{tab:Compact lattice models and design results}, the $SR$s were relatively high for small 2D HP models, but they decreased as $N$ increased. Nevertheless, in the case of $N = 16$, the $SR$ is higher than that for the smaller case, $N =12$. The $N = 16$ case has an extremely high percentage of non-maximally compact conformations than the $N = 12$ one (Table I). Hence, this result shows that the proposed design method is more efficient for non-maximally compact conformations.

The average degeneracy $d_{{\rm av}}$ was low for 2D models. By contrast, the success rate for 3D models 
\begin{table}[]
	\begin{center}
		\caption{Design results and the optimal chemical potential $\mu^{*}$}
			\begin{tabular}{ccccccc|cc}\hline \hline
			Size&$N_{\rm c}^{(\rm g)}$ & $N_{\rm c}^{(\rm m)}$ & $N_{\rm c}^{(\rm b)}$ & $SR$ &$d_{{\rm av}}$& $\mu^{*}$&$(\epsilon_{1}, \epsilon_{2}, \epsilon_{3})$\\[1.5pt] \hline
			\hspace{0.5mm}$N = 9$ & 7 & 1 & 0 & 87.5 &1.25& $0.55$& \\
			\hspace{0.5mm}$N = 12$ & 29 &11&0& $72.5$ & 1.475& 0.6& \\
			\hspace{0.5mm}$N = 16$ & 393&89&0 & 81.5 & 1.26&0.62 &$(-1,0,0)$\\
			\hspace{0.5mm}$5\times5$ & 68 & 32 & 0 & 68 &1.48&$0.74$ &\\
			\hspace{0.5mm}$6\times6$ & 63 &37&0& $63$ &1.58& 0.8 &\\ \cline{8-8}
			$2\times2\times3$ & $17$ &30&1& 35.4 &2.94& 1.7 &\multirow{2}{*}{$(-2.3,-1,0)$}&\\ 
			$3\times3\times3$ & 8 & 80 & 12 & 8& 10.67&2.33 &\\ \hline \hline
			\end{tabular}
			\label{tab:Compact lattice models and design results}
	\end{center}
\end{table}
was low compared with that of 2D models. For $2\times2\times3$, $d_{{\rm av}}$ was low, but for $3\times3\times3$, it was comparatively high. Thus, designed sequences did not appear to be likely to fold into the target conformations for the $3\times3\times3$ cubic lattice. In addition, $\mu^{*}$ increased as the number of residues increased for both the 2D and 3D lattices.
 
Nevertheless, to design the 3D lattice conformations of the HP model efficiently is difficult because the logarithm of the number of types of the amino acid (alphabet size) is smaller than the conformational entropy of a residue\cite[Sec.IV]{pande2000heteropolymer}. Therefore, for $N = 2 \times 2 \times 3$ and $3 \times 3 \times 3$, design accuracy would be low when using the HP model. We thus show the design results by increasing the alphabet size for the 3D lattice cases in the next subsection.

\begin{figure}[b]
	\begin{center}
		\includegraphics[width=8.5cm]{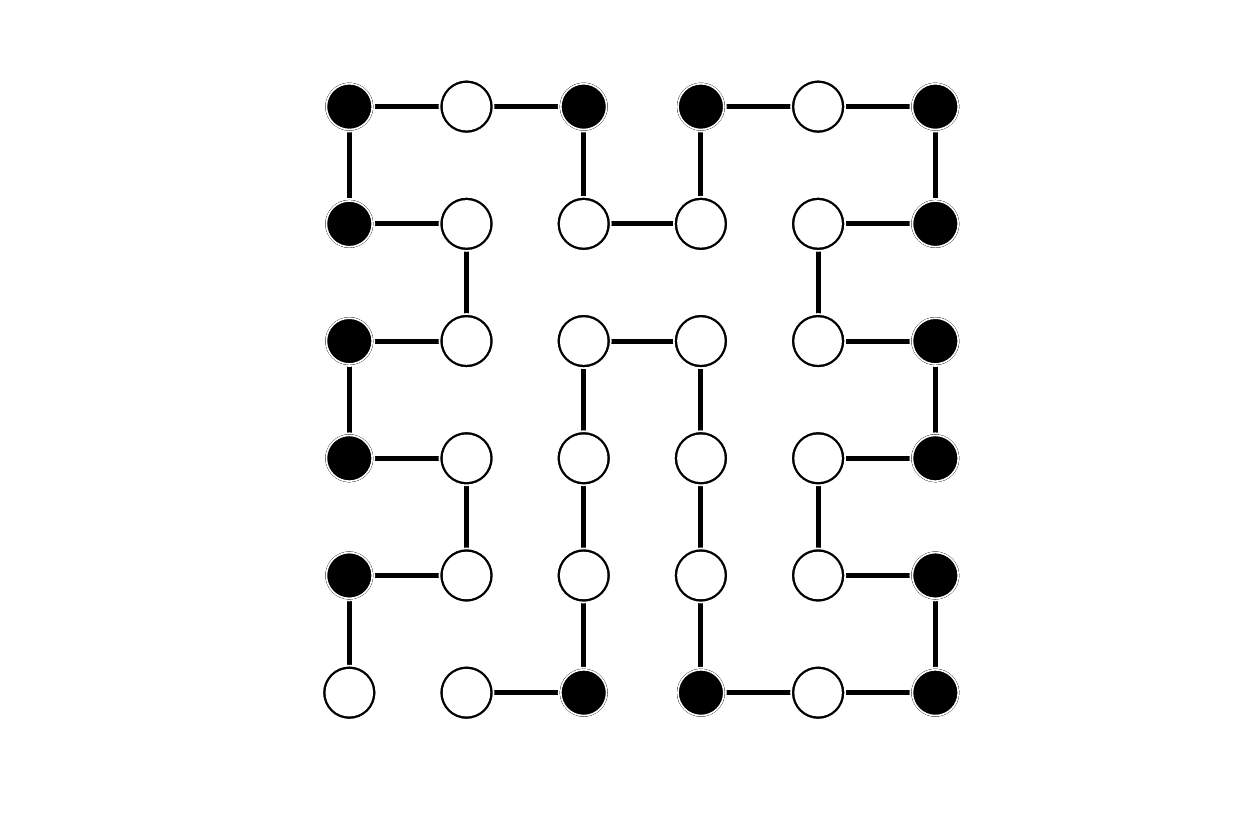}
	\end{center}
	\caption{Designed sequence of the MHDC of $6\times6$ HP model with $(\epsilon_{1}, \epsilon_{2}, \epsilon_{3}) = (-1,0,0)$, $\beta = 10$, and $\mu^{*} = 0.8$. The white and black balls denote H and P residues, respectively (the same applies in the following figures.}
	\label{fig:6_6_MHDC}
\end{figure}
\begin{figure}
	\begin{center}
			\includegraphics[width=8.5cm]{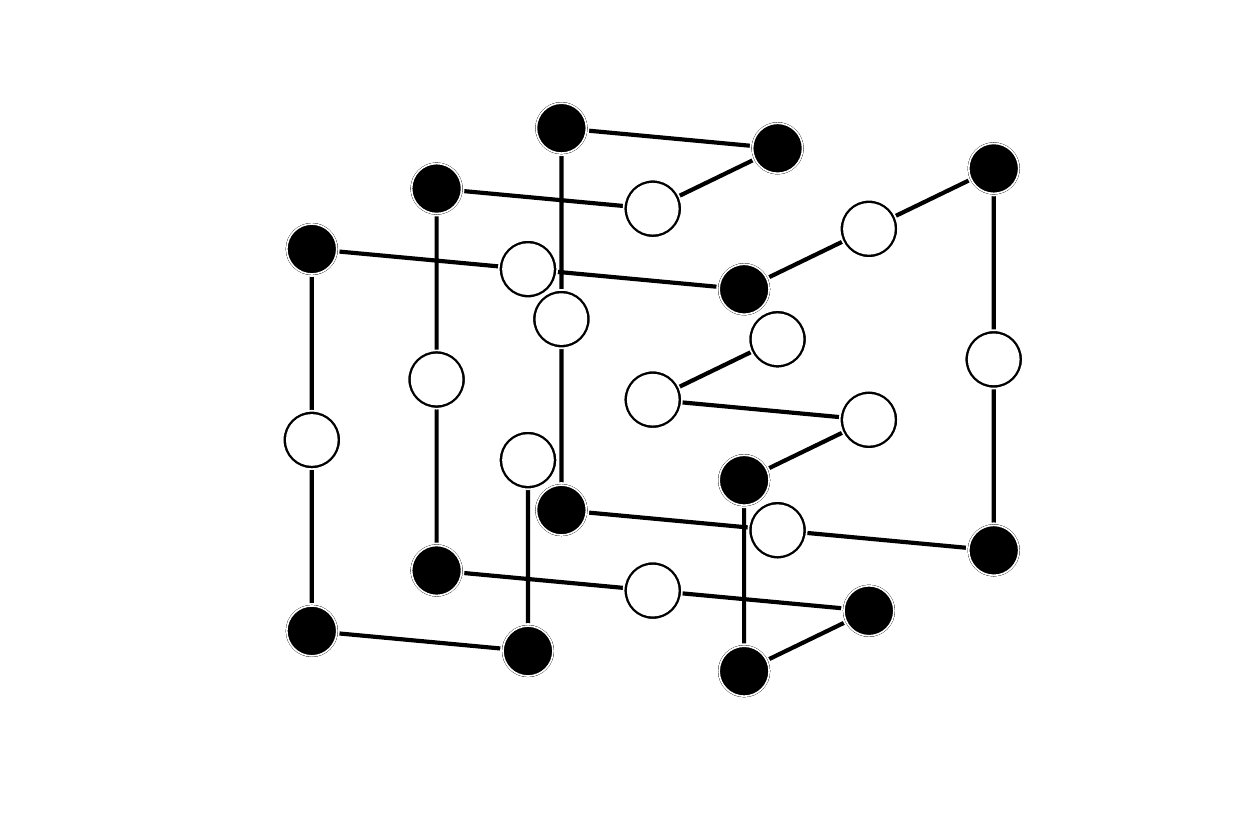}
	\end{center}
	\caption{	Designed sequence of the MHDC $3\times3\times3$ HP model with $(\epsilon_{1}, \epsilon_{2}, \epsilon_{3}) = (-2.3,-1,0)$, $\beta = 10$, and $\mu^{*} = 2.33$.}
	\label{fig:3_3_3_MHDC}
\end{figure}
\begin{figure}
	\begin{center}
		\includegraphics[width=8.5cm]{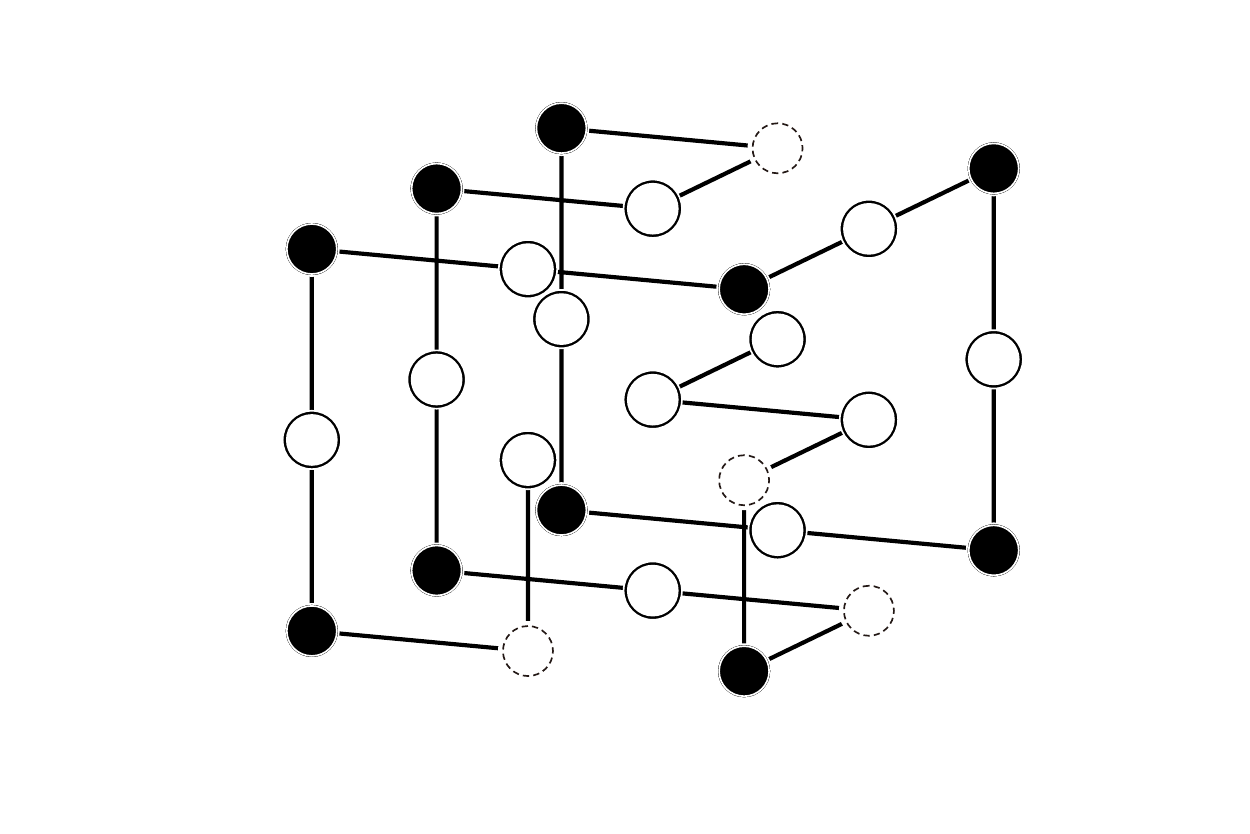}
	\end{center}
	\caption{
	Designed sequence of the MHDC of $3\times3\times3$ HP model with $(\epsilon_{1}, \epsilon_{2}, \epsilon_{3}) = (-1,0,0)$, $\beta = 10$, and $\mu^{*} = 1.0$.}
	\label{fig:3_3_3_MHDC_-1_0_0}
\end{figure}

Note that we did not enumerate designabilities of all conformations for the $5\times5$, $6\times6$, and $3\times3\times3$ models; hence, there may have been non-designable conformations among the 100 randomly chosen conformations.  However, it is likely that this was not the case for the $5\times5$ and $6\times6$ models, because the smaller \erase{$3\times4$} $N = 9$ HP model did not lead to any non-designable conformation. On the other hand, the fraction of non-designable conformations out of all conformations for the $2\times2\times3$ model was 21/69; the fraction for the $3\times3\times3$ model is expected to be less than that because the fraction decreased as the size increased in the 2D cases. Thus, there may have been a considerable number of non-designable conformations among the randomly chosen 100 conformations for the $3\times3\times3$ model; hence, the real success rate of the $3\times3\times3$ model increased when non-designable conformations were excluded.

Concerning the MHDC of the $6\times6$ and $3\times3\times3$ HP models, we obtained a good sequence (Figs. \ref{fig:6_6_MHDC}, \ref{fig:3_3_3_MHDC}, and \ref{fig:3_3_3_MHDC_-1_0_0}). This is the first example of design of a $6\times6$ MHDC without enumerating all HP sequences\cite{li1996emergence}.
For the MHDC of the $3\times3\times3$ HP model, we successfully designed a good sequence for the energy parameters $(\epsilon_{1}, \epsilon_{2}, \epsilon_{3}) = (-2.3,-1,\,0)$ (Fig. \ref{fig:3_3_3_MHDC}) and (-1, 0, 0) (Fig. \ref{fig:3_3_3_MHDC_-1_0_0}). We executed $10^4$ MCSs for these three cases. The results obtained here demonstrate the features of general globular proteins, with H residues on the inside of the protein and P residues on the surface exposed to the surrounding water molecules. We observed four residues (surrounded by dotted black circles in Figs. \ref{fig:3_3_3_MHDC} and \ref{fig:3_3_3_MHDC_-1_0_0}) that were different from each other, possibly owing to the presence or absence of H-P (P-H) contact energies.

The larger $\mu^{*}$ of the MHDC of $3\times3\times3$ HP model with $(\epsilon_{1}, \epsilon_{2}, \epsilon_{3}) = (-2.3,-1,\,0)$ compared with the case of $(\epsilon_{1}, \epsilon_{2}, \epsilon_{3}) = (-1,0,0)$ could have been due to the lower H-H interaction $\epsilon_{1} = -2.3$, leading to a greater increase in the number of H-residues than in the case of $(\epsilon_{1}, \epsilon_{2}, \epsilon_{3}) = (-1,0,0)$. Therefore, one needs to increase $\mu^{*}$ in order for the surface residues to be P residues.

\subsection{Enumerable 3D conformations with increasing alphabet size}

In the results shown in Table II, the $SR$s of 3D cases are quite low. As mentioned above, this is because the logarithm of the alphabet size is smaller than the conformational entropy of a residue in the case of 3D lattice models\cite[Sec.IV]{pande2000heteropolymer}.

\begin{table}[]
	\begin{center}
		\caption{Design results of small 3D compact lattice conformations with Miyazawa-Jernigan matrix\cite{miyazawa1985estimation} and the optimal chemical potential $\mu^{*}$ with the results of HP model in Table. \ref{tab:Compact lattice models and design results}}
			\begin{tabular}{ccccccccc}\hline \hline
			Size&Alphabet size&$N_{\rm c}^{(\rm g)}$ & $N_{\rm c}^{(\rm m)}$ & $N_{\rm c}^{(\rm b)}$ & $SR$ &$d_{{\rm av}}$& $\mu^{*}$&\\[1.5pt] \hline
			\multirow{2}{*}{$2\times2\times3$}&20 (MJ)& 16 &18& 14& 33.3&1.60& 1.1 \\ 
										& 2 (HP) & 17 &30& 1& 35.4&2.97& 1.7 \\ 
			 \multirow{2}{*}{$3\times3\times3$}& 20 (MJ)& 19 & 63 & 18 & 19& 5.74&1.55 \\ 
							 &2 (HP)& 8 & 80 & 12 & 8& 10.67&2.33 \\ \hline \hline
			\end{tabular}
			\label{tab:3D with MJ 85}
	\end{center}
\end{table}

Therefore, we show the design results of the 3D lattice conformations with increasing alphabet size here. The 3D lattice target conformations designed here are the same as those given in the previous subsection. The alphabet size is 20, and we use the original Miyazawa-Jernigan (MJ) matrix\cite[upper half of Table V]{miyazawa1985estimation} for the contact energy of all the pairs of amino acids. For simplicity, we set all interactions between water molecules and polar amino acids to be equal. The procedure of optimizing $\mu$ is identical to the case that of the HP model. We assume that the amino acids Y, F, W, L, V. I, A, P, and M are hydrophobic\cite{monera1995relashonship}.

It is impossible to calculate the expectation value $\left<\sigma_{i}\right>$ in the same way as the HP model because the 20 types of amino acids cannot be represented using the Ising variables. The optimal $\left<\sigma_{i}\right>$, therefore, is given by the type sampled the most after the burn-in period.

\begin{figure}[b]
	\begin{center}
		\includegraphics[width=8.5cm]{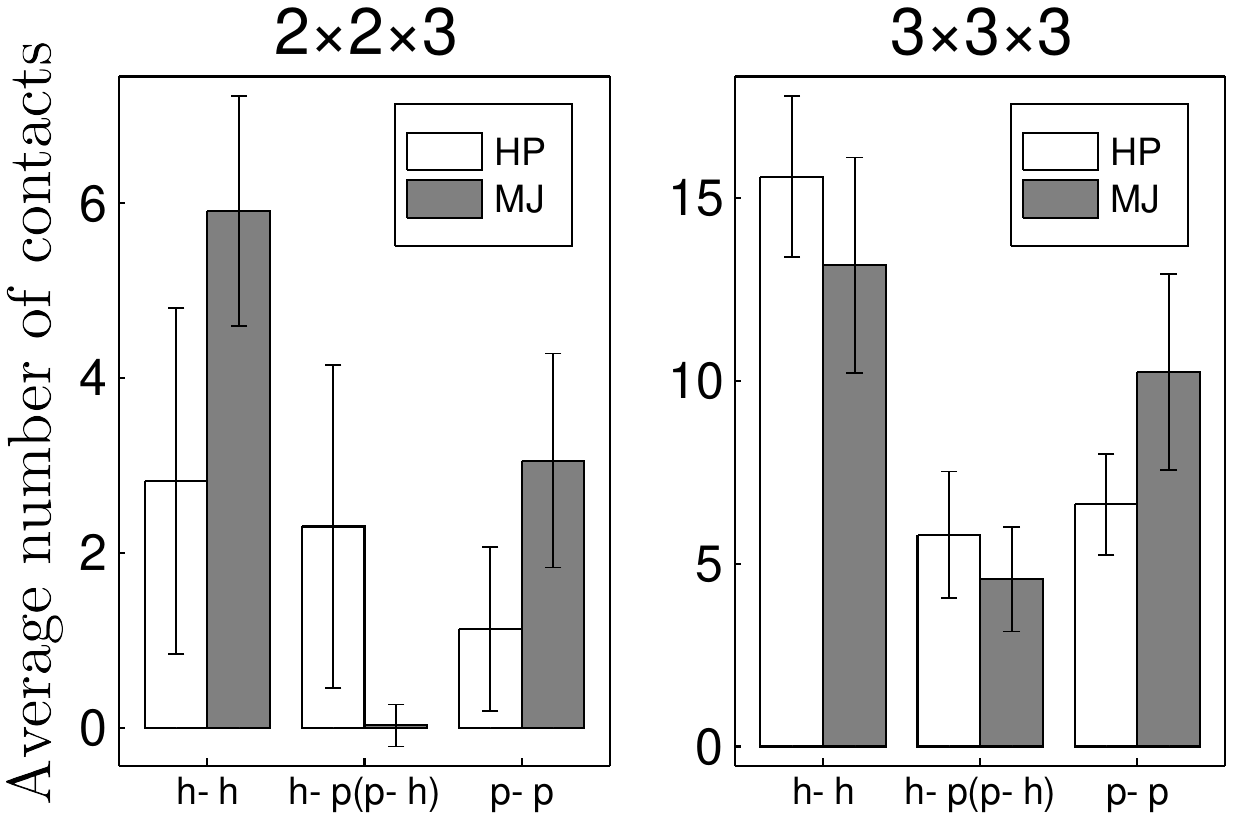}
	\end{center}
	\caption{
	The number of contacts for the three types, hydrophobic-hydrophobic (h-h), hydrophobic-polar (polar-hydrophobic) (h-p, p-h), and polar-polar (p-p) contacts, for the two types of energy parameters, the HP model with $(\epsilon_{1}, \epsilon_{2}, \epsilon_{3}) = (-2.3,-1,0)$ and MJ. The value of each bar is an average of the number of contacts for each type of all the designed conformations. The each error bar represents a standard deviation. The total number of contacts is 9 for $N = 2\times2\times3$, and 28 for $3\times3\times3$.
	}
\end{figure}

Table III depicts the obtained results.  To obtain the precise results, one has to calculate the designablities of all conformations using the MJ matrix. This is computationally difficult because calculating the $20^{N}$ energy patterns for all conformations is necessary. Nevertheless, the designability of the 20-letter model correlates with the designability of the 2-letter model\cite{li2002designability}. Hence, in the case of $N = 2\times2\times3$, as given in Table II described in previous subsection, we excluded 21 non-designable conformations when we assumed the energy parameter $(\epsilon_{1}, \epsilon_{2}, \epsilon_{3}) = (-2.3, -1.0, 0)$ of the HP model. 

According to results summarized in Table III, for $N = 2\times2\times3$, the value of $SR$ of the 20-letter is a slightly less than the $SR$ of the HP model. By contrast, in the case of $N = 3\times3\times3$, $SR$ of the 20-letter is more than twice as large as the $SR$ of the 2-letter. The ground state degeneracy typically breaks upon increasing the alphabet size.  We believe that an increase in $SR$ for $N = 3\times3\times3$ is a result of the aforementioned degeneracy breaking.  The difference in the changing of $SR$ between the above two cases is, we consider, because of the presence or absence of the core residue. 

We discuss how the presence or absence of the core residue affects the above difference. In Fig. 5, we represent the change in the number of contacts for the three types, hydrophobic-hydrophobic, hydrophobic-polar (polar-hydrophobic), and polar-polar contacts, when the energy parameter changes from (-2.3, -1, 0) (HP model) to the MJ matrix. However, in the case of $N = 2\times2\times3$, the number of hydrophobic - polar contacts almost vanishes while changing the energy parameter from (-2.3, -1, 0) (HP model) to the MJ matrix, as shown in Fig. 5. In. contrast, in the case of $N = 3\times3\times3$, even though the number of polar-polar contacts also increases, the balance among the distribution of the three types of contacts does not significantly change.  We consider that this difference in the distribution of the three types of contacts shown in Fig. 5 is the reason for a difference in the change in $SR$ between the two cases.  In the MJ matrix\cite[upper half of Table V]{miyazawa1985estimation}, contact energy increases in the following order: of hydrophobic-hydrophobic, hydrophobic-polar (polar-hydrophobic), and polar-polar contacts. Thus, using the MJ matrix, a conformation with no hydrophobic-polar contacts rarely becomes a ground state. The value of $SR$ for $N = 2\times2\times3$ is almost constant because this vanishing of the polar-polar contacts and the degeneracy breaking offset each other.

In addition, $N_{\rm c}^{(\rm b)}$ (the number of conformations designed by bad sequences) increase upon increasing the alphabet size in the cases of both $N = 2\times2\times3$ and $3\times3\times3$. The degeneracy breaking affects this result as well as increases $N_{\rm c}^{(\rm g)}$. Additionally, one of the reasons for an increase in $N_{\rm c}^{(\rm b)}$ is that the designabilities of the 20-letter model are less than the designabilities of the 2-letter model for many conformations\cite{li2002designability}.

The average degeneracies $d_{{\rm av}}$ of the two cases decrease with increasing the alphabet size. Thus, this is an evidence of the degeneracy breaking. 

\subsection{Comparison to previous method}

In this subsection, using the lattice HP model, we present the results of a comparison between our Bayesian design method and a conventional method involving the exhaustive conformational search. We choose the dual MC method by Seno {\it et al.}\cite{seno1996optimal} as aforementioned conventional design method.  

\begin{table*}
	\begin{center}
		\caption{
			We carried out the exact calculations of $Z(\boldsymbol \sigma)$ for the dual MC method because the target conformations are all enumerable. We designated the name the dual MC method for the sake of convenience. For the dual MC method, the temperature of $k$-th MC step is given by $T_{k} = T_{0}/(1 + \alpha k^{2})$ where $T_{0}$, $k$, $T_{k}$, and $\alpha>0$ denotes the initial temperature, MC step, temperature of a MC step $k$, and the controlling parameter respectively. For all the lattice protein sizes, the initial temperature is represented by $T_{0} = 10$, and $\alpha$ is the value of the above equation in which we substitute the terminal temperature $T = 0.1$ for $T_{k}$ and the MCSs for $k$. DT and DE denote the design time and the design efficiency, respectively. The MCSs of the Bayesian method are set as low as possible to achieve the same $SR$ value as reported in Table II as described in the main text. The last MCS, which is mentioned in the last column of the dual MC method, denotes the average last MC step for all conformations. All MCSs values for the dual MC method range from 500 to 70000.  These MCSs differ depending on the lattice protein size as well as the conformations of the same size. 		
		}
		\begin{tabular*}{17.5cm}{@{\extracolsep{\fill}} ccccc|cccc} \hline \hline
			& \multicolumn{4}{c|}{Bayesian method} & \multicolumn{4}{c}{Dual Monte Carlo method} \\
			Size & SR (\%)& DT (s) & DE (\%/s) & MCSs & SR (\%) & DT (s) & DE (\%/s)&  Last MCS \\ \cline{1-2} \hline
			$N = 9$ &  87.50 & 0.02615& 3103 & 1000&  100.0 & 1.043& 96.12 & 492\\
			$N = 12$ & 72.50 & 0.04356& 1670 &2000 &90.00 & 4.300& 23.53 & 2509\\
			$N = 16$ &81.53 & 0.1422& 5708 & 5000&92.74  & 167.6 & 0.7182 & 17924 \\
			$5 \times 5$ & 68.00&  1.179& 577.3 & 30000 & 92.00 & 477.5 & 0.2281 & 19842\\
			$2 \times 2 \times 3$ &35.42 &  0.2513& 136.9 &10000 & 85.41 & 40.77& 6.407 & 10169 \\ \hline \hline
		\end{tabular*}
		\label{tab:comparison data}
	\end{center}
\end{table*}

The dual MC method maximizes the target probability as given by Eq. (\ref{CanonicalDist}); hence, one can intuitively predict that the $SR$ of the dual MC method is larger.  The dual MC method calculates $Z_{\beta}(\boldsymbol \sigma)$, which corresponds to the exhaustive conformational search; therefore, the calculation time becomes typically longer.

We compare the $SR$, calculation time (design time), and the $SR$ per design time (design efficiency).  Using design methods based on the MTP criterion, one can solve the design problem if design time is as long as possible.  Therefore, comparing only the $SR$ is unreasonable if the design time is not limited. Therefore, in this study, we compare the $SR$ as well as the design efficiency. This viewpoint is especially significant for real applications.

The enumerable conformations shown in Table I are the ones which were designed for this comparison.  Nevertheless, we did not design the 2D $N = 6\times6$ and the 3D $N = 3\times3\times3$ conformations except for the MHDC of $3\times3\times3$, because their sequence spaces are extremely large to be be  designed using the dual MC method.  The design time of a conformation of $N = 6\times6$ and $3\times3\times3$ is about 10 -- 30 days using a normal PC (1.2 GHz dual-core Intel Core m3 and 8 GB memory) by our estimation.  Note that we carried out the exact calculation of $Z(\boldsymbol \sigma)$ for the dual MC method because all the target conformations are enumerable. Thus, we carried out the MC sampling (simulated annealing) of the sequence spaces only (therefore, we name this method the dual MC method for convenience).

The conditions for the design calculations are as follows. For our Bayesian method, for each size, $\mu^{*}$ is identical to the value in Table \ref{tab:Compact lattice models and design results}, and the number of MCSs is set as low as possible to achieve the same $SR$ as reported in Table II. For the simulated annealing of the dual MC method, the terminal temperature is $T = 0.1$; hence, it equals the terminal temperature of our Bayesian method, $\beta = 10$. The cooling schedule of $T$ is not the linear function used in \cite{seno1996optimal} but is an inverse power function of the Monte Carlo step because the latter function avoids getting trapped into a metastable state in the amino acid sequence space as compared to the linear function (we show the equation of this cooling schedule in the caption of Table IV below).

As shown in Table IV, for all the cases, the $SR$s of the dual MC method surpass the $SR$s of the Bayesian method, especially for the 3D $N = 2\times2\times3$. On the contrary, the DTs of the Bayesian method are all significantly less than the DTs of the dual MC method: the former ones are of the order of 1/100 to 1/1000 of the latter ones.  Thus, each DE of the Bayesian method is about 100 or 1000 times each DT in the of the dual MC method.  Our Bayesian method is quite efficient compared to the dual MC method.  Furthermore, the DT of the MHDC of the 3D $N = 3\times3\times3$ using the Bayesian method is 0.9244s, but it is about 434600s (about five days) for the dual MC method.  It indicates that the DTs' difference between the Bayesian method and the design methods that have $Z(\boldsymbol \sigma)$ increases as the number of residues increases.

\subsection{Large 2D conformations}
\begin{figure}[h]
		\includegraphics[width=8.5cm]{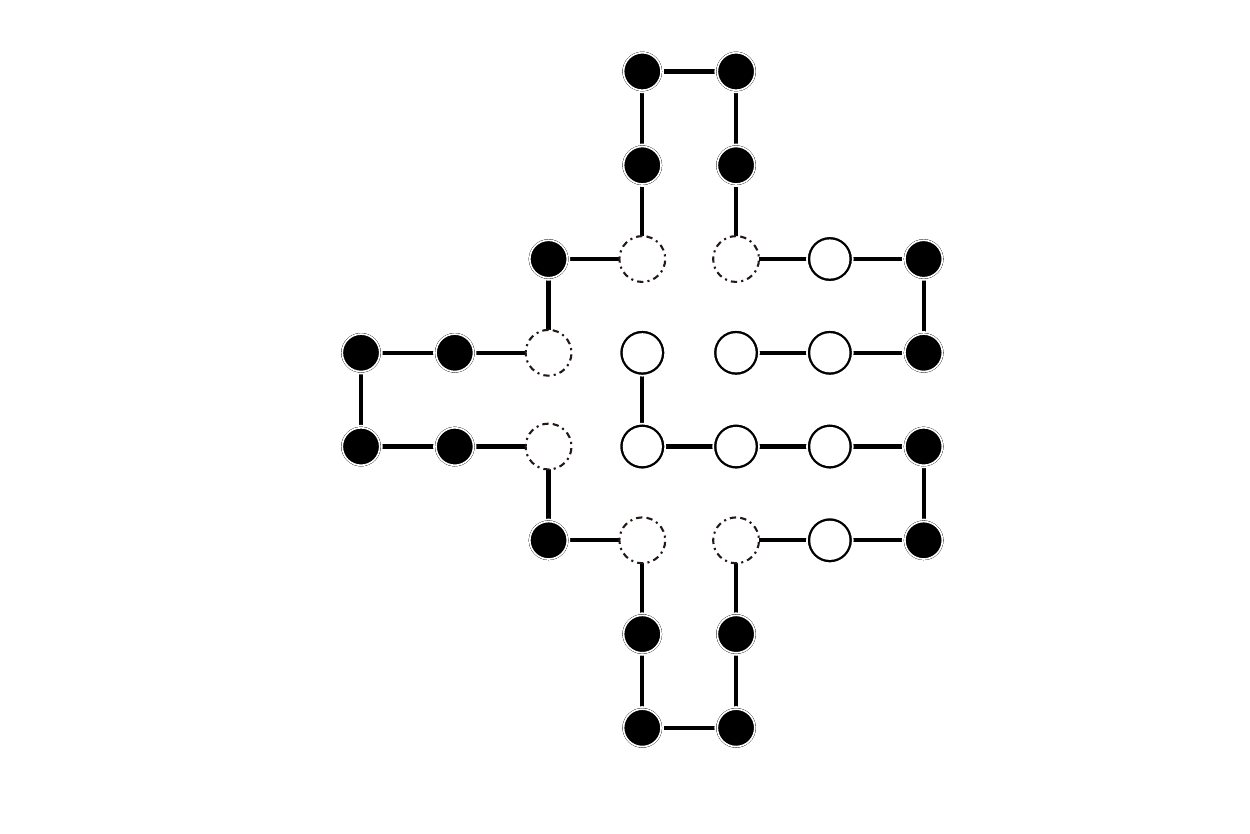}
		\caption{
		Designed sequence of the $N = 32$ 2D HP model with $(\epsilon_{1}, \epsilon_{2}, \epsilon_{3}) = (-1,0,0)$, $\beta = 10$, and $\mu^{*} = 0.7$.
		}
		\label{fig:Irback_N_32}
\end{figure}

\begin{figure}[h]
	\begin{center}
		\includegraphics[width=8.5cm]{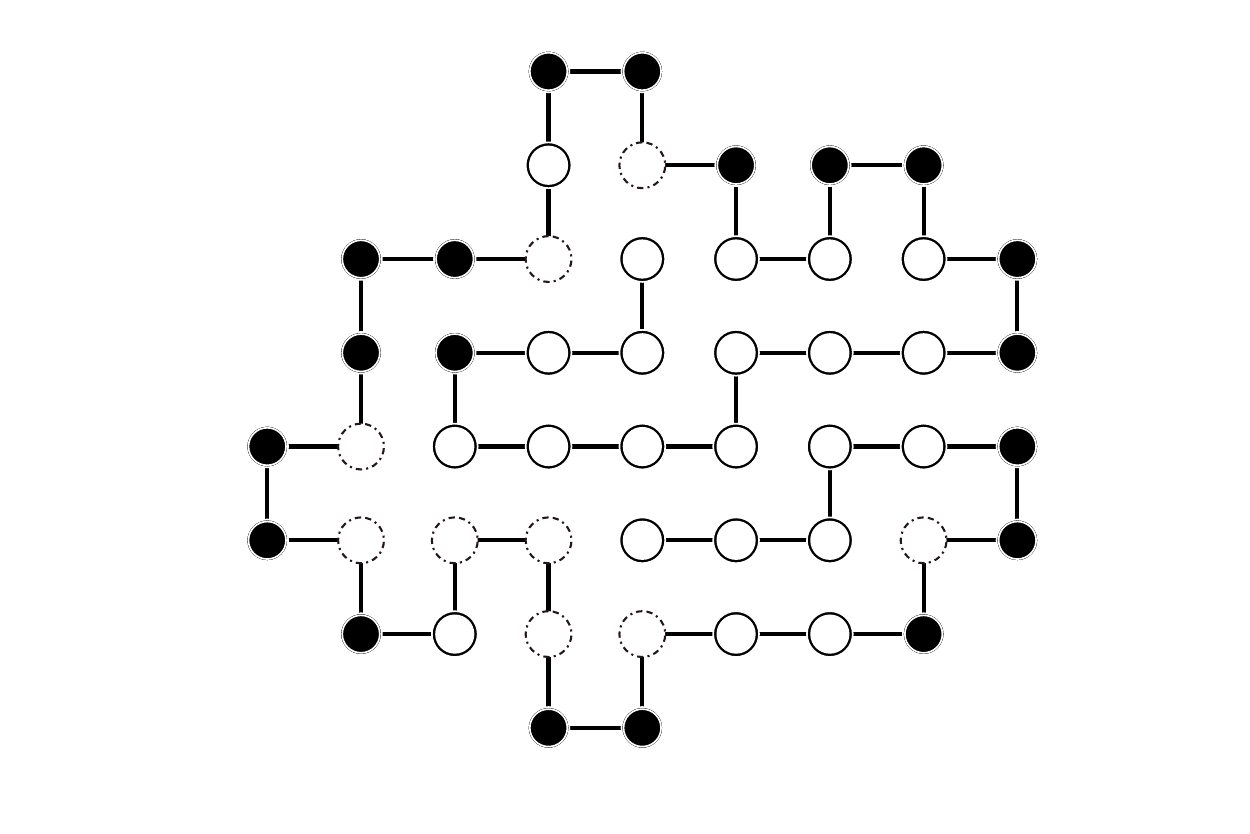}
		\caption{
		Designed sequence of the $N = 50$ 2D HP model with $(\epsilon_{1}, \epsilon_{2}, \epsilon_{3}) = (-1,0,0)$, $\beta = 10$, and $\mu^{*} = 0.85$.
		}
		\label{fig:Irback_N_50}
	\end{center}
\end{figure}

Here, we chose 2D HP models with comparatively large size ($N = 32, 50$) models studied by Irb\"{a}ck {\it et al.}\cite{irback1998monte, irback1999design}. This confirmed that the designed sequence was likely to fold into the target conformation with simulated tempering. For the model with $N = 32$ (respectively 50), the parameters were set to $\mu^{*} = 0.7$ (0.85) and the MCSs were $10^4$ ($10^5$). The energy parameters were set to $(\epsilon_{1}, \epsilon_{2}, \epsilon_{3}) = (-1,0,0)$ in both cases. The simulation was executed by a normal PC with 1.2 GHz dual-core Intel Core m3 and 8 GB memory, and the calculation time was approximately 0.5--1 s (11--12 s) for $N=32$ (50). Thus, our method ran faster than those used in previous studies. As a result, we successfully designed the same sequences reported by Irb\"{a}ck {\it et al.} (Figs. \ref{fig:Irback_N_32} and \ref{fig:Irback_N_50}). Our method also demonstrates the features of globular proteins.

\begin{figure}[t]
	\begin{center}
		\includegraphics[width=8.0cm]{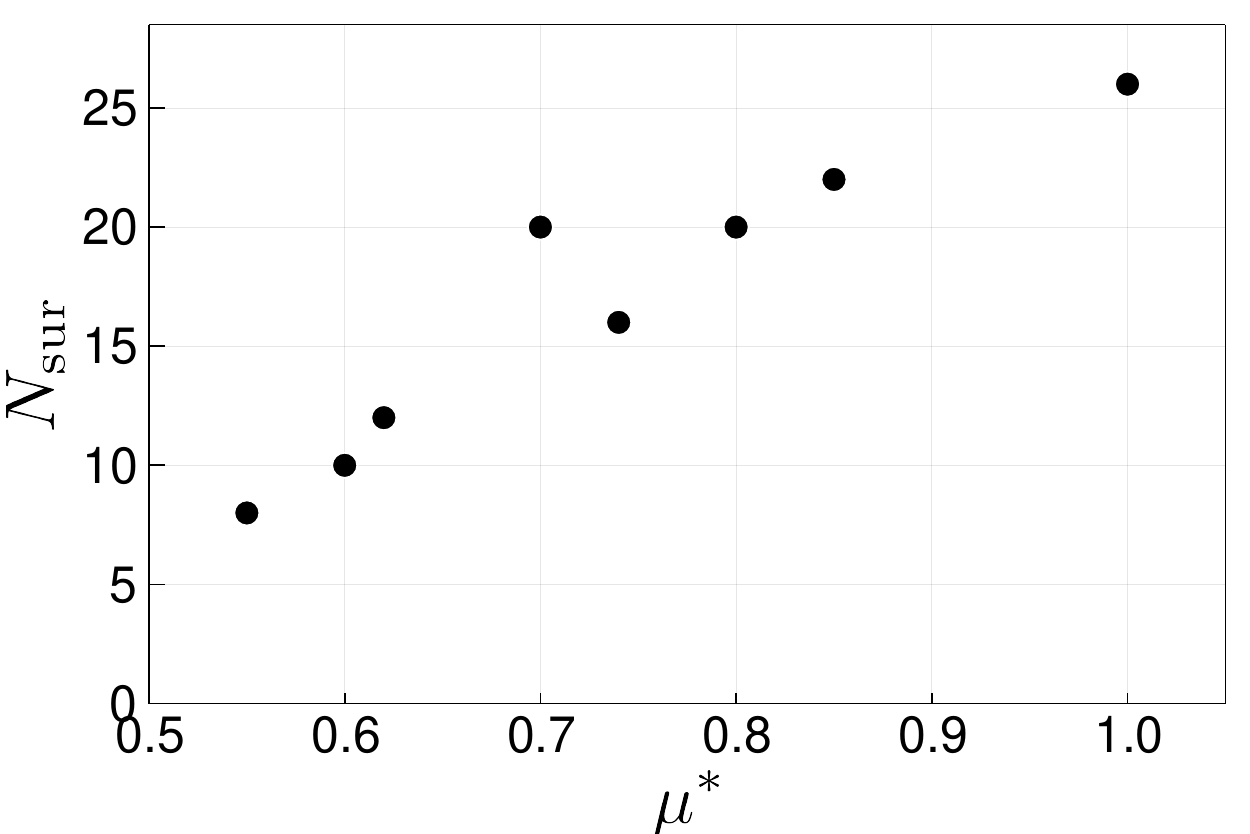}
		\caption{Relation between $\mu^{*}$ and number of surface residues, $N_{\rm sur}$.}
		\label{fig:mu_vs_surface}
	\end{center}
\end{figure}

\subsection{Optimal $\mu^{*}$ and number of surface residues}
We represent the relation between the optimal $\mu^{*}$ and the number of surface residues $N_{\rm sur}$ in Fig. \ref{fig:mu_vs_surface}. We show only the results for $(\epsilon_{1}, \epsilon_{2}, \epsilon_{3}) = (-1,0,0)$ because the optimal $\mu^{*}$ varies depending on the energy parameters for a given conformation. We therefore plotted the results for all 2D models and the $3\times3\times3$ MHDC model with $(\epsilon_{1}, \epsilon_{2}, \epsilon_{3}) = (-1,0,0)$ The residues that were bent 90 degrees inward (indicated by a dashed black circle in Figs. \ref{fig:Irback_N_32} and \ref{fig:Irback_N_50}) were not counted for $N_{\rm sur}$ because a water molecule is unlikely to combine with such residues (see Fig. \ref{fig:mu_vs_surface}).

\vspace{-1.0mm}
We observed noticeable linearity between $\mu^{*}$ and $N_{\rm sur}$. The outlier $(\mu^{*}, N_{\rm sur}) = (0.70, 20)$ was obtained in the 2D $N = 32$ case (Fig. \ref{fig:Irback_N_32}), in which the target conformation was not fully compact and the number of surface residues was much larger than those of other target conformations tested. According to these results, the optimal $\mu^*$ can be estimated by the number of surface residues of a target conformation.

\subsection{Probability of a P residue}
\begin{figure}
	\begin{center}
	\includegraphics[width=8.5cm]{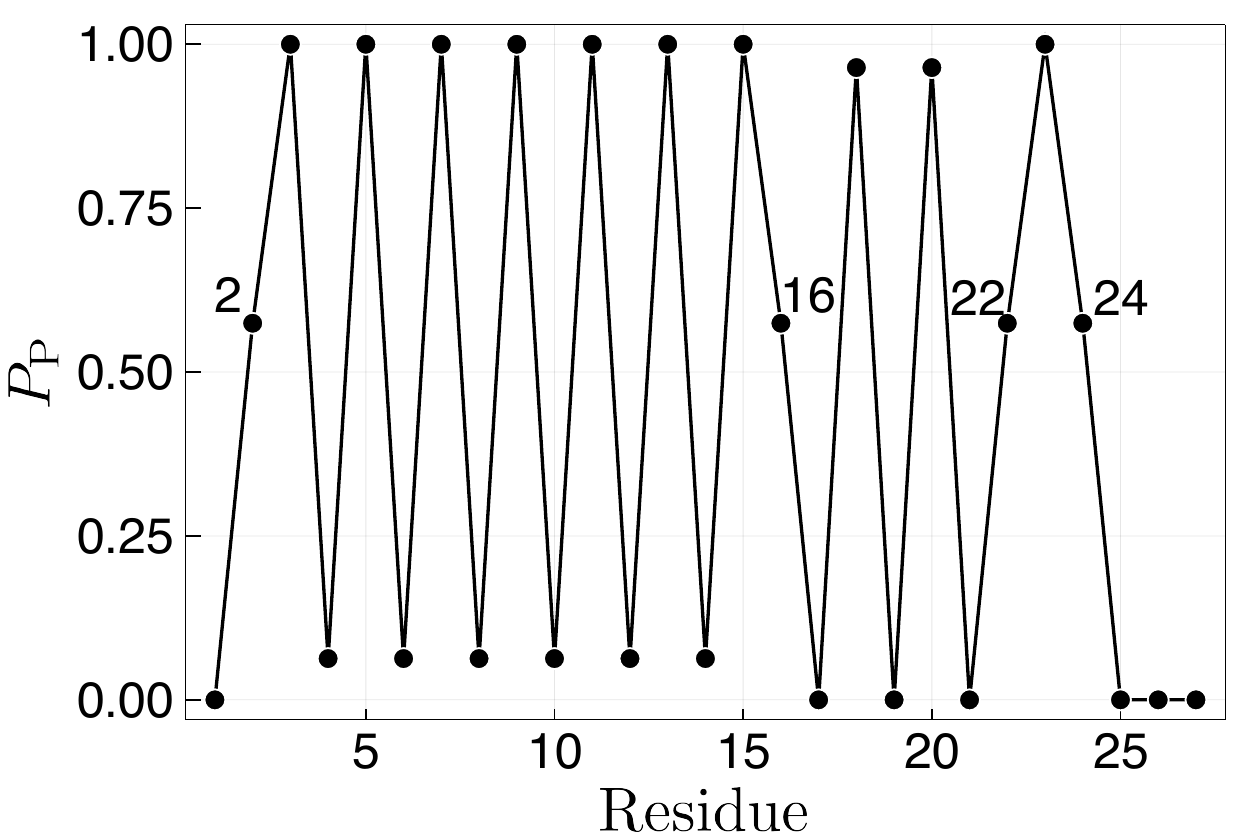}
	\caption{$P_{\rm P}$ of each residue of $3\times3\times3$ MHDC with $(\epsilon_{1}, \epsilon_{2}, \epsilon_{3}) = (-2.3,-1,\,0)$ and $\mu = 2.33$ (Fig. \ref{fig:3_3_3_MHDC}). The residue number starts from the center residue of the front side of Fig. \ref{fig:3_3_3_MHDC}.}
		\label{fig:Pp_3_3_3_MHDC}
	\end{center}
\end{figure}

\begin{figure}
	\begin{center}
	\includegraphics[width=8.5cm]{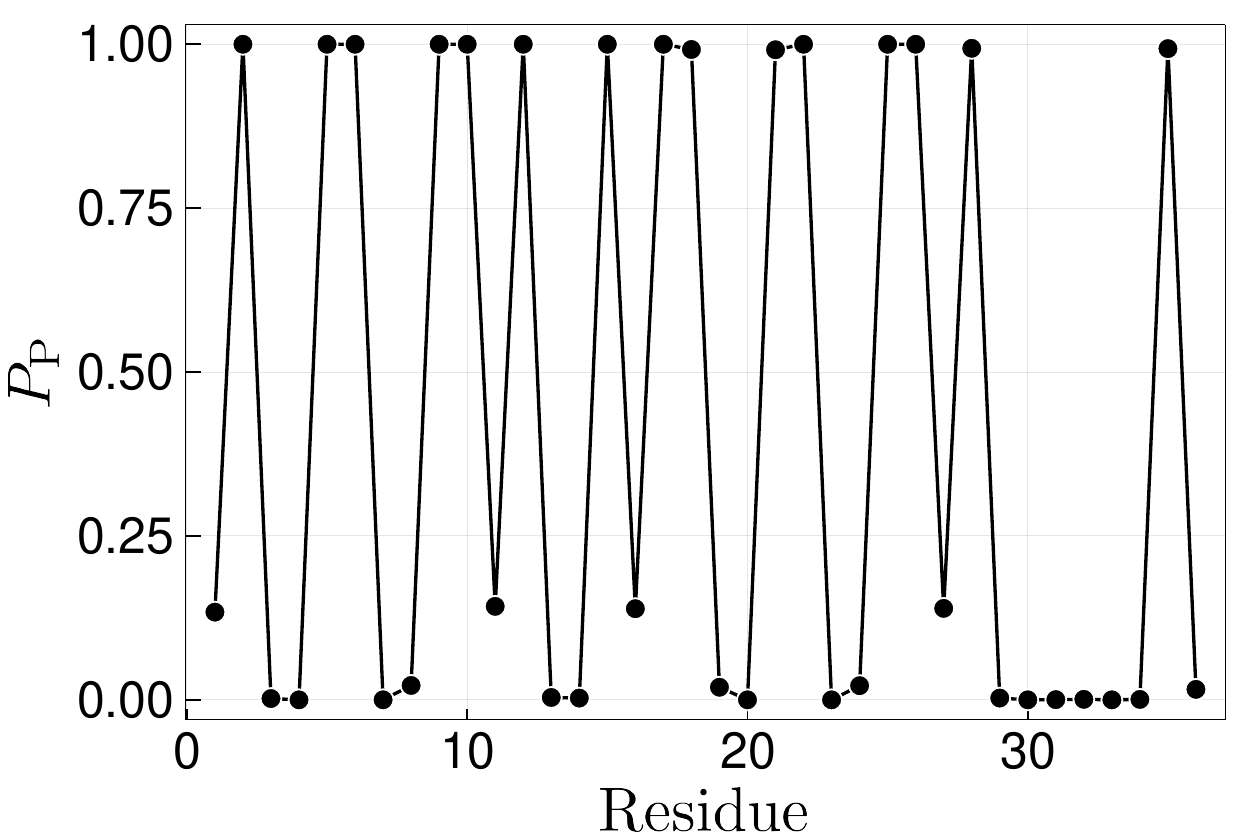}
	\caption{$P_{\rm P}$ of each residue of $6\times 6$ MHDC with $(\epsilon_{1}, \epsilon_{2}, \epsilon_{3}) = (-1,0,0)$ and $\mu = 0.8$ (Fig. \ref{fig:6_6_MHDC}). The residue number starts from the bottom left residue of Fig. \ref{fig:6_6_MHDC}.}
	\label{fig:Pp_6_6_MHDC}
	\end{center}
\end{figure}

Finally, in order to clarify why 3D models performed less well than 2D models, 
we consider the probability $P_{\rm P}$ that a residue is P for all residues of the $3\times3\times3$ and $6\times6$ MHDC models (Figs. \ref{fig:Pp_3_3_3_MHDC} and \ref{fig:Pp_6_6_MHDC}). We use $p(\sigma_{i}=-1|\boldsymbol R ; \boldsymbol \sigma_{\backslash i})$ in Eq. (\ref{GSProb}) as $P_{\rm P}$. 
Each $P_{\rm P}$ in Figs. \ref{fig:Pp_3_3_3_MHDC} and \ref{fig:Pp_6_6_MHDC} is the average of $1/\left(1+\exp\left[-{\beta}\{\Delta E_{i}(\boldsymbol R; \boldsymbol \sigma) + \mu\}\right]\right)$ over the last $1/5$ MCSs in both cases. In the case of the 3D $3\times3\times3$ lattice, the values of $P_{\rm P}$ for residues 2, 16, 22, and 24 were not very high. These residues were located in the center of each cube side. By contrast, all $P_{\rm P}$ values greater than 0.5 were almost equal to 1 in the case of the 2D $6\times6$ lattice. Accordingly, it can be seen that the clear division of all $P_{\rm P}$ values into $1$ or $0$ is an index of successful design.
Thus, the 3D models are difficult instances for our design method.

\section{DISCUSSION}
Our method is similar to the SG method proposed by Shakhnovich and Gutin\cite{shakhnovich1993engineering}, which did not include $Z(\boldsymbol \sigma)$.  The difference between the SG method and ours is the minimization function. The SG method minimizes $E(\boldsymbol R ;\boldsymbol \sigma)$ directly, keeping $N_{\rm P}(\boldsymbol \sigma)$ a constant value determined {\it a priori}, but our method minimizes $E(\boldsymbol R ; \boldsymbol \sigma) - \mu N_{\rm P}(\boldsymbol \sigma)$. Therefore, our method can minimize $E(\boldsymbol R ;\boldsymbol \sigma)$, maintaining the general features of globular proteins, that is, H residues on the inside and P residues on the surface exposed to the surrounding water molecules. Thus, one can minimize $E(\boldsymbol R ;\boldsymbol \sigma)$ while reducing the diversity of conformations into which a designed HP sequence can fold by minimizing $E(\boldsymbol R ; \boldsymbol \sigma) - \mu N_{\rm P}(\boldsymbol \sigma)$. This corresponds, in a sense, to negative design\cite{jin2003novo}.

On the other hand, as discussed above, our method failed in the cases of 3D HP models and comparatively large compact 2D HP models because it failed to reduce the diversity of the foldable conformations of the designed sequence in these cases.  The diversity---or, more simply, the total number of self-avoiding walks---increases for 3D lattices compared with 2D lattices. Such diversity is expected to increase as $N$ increases even in the case of 2D lattices. Thus, the success rates decrease in these cases.

In addition, the numbers of core residues in the 3D models used in this study were low, e.g., the 3D $2\times2\times3$ model had no core residue and the 3D $3\times3\times3$ model had only one; hence, it was difficult to design globular protein-like conformations using these models. Given that our design method finds an optimal sequence by controlling $\mu$,  such small numbers of cores may explain the lower performance of our method in the case of the 3D models. If our design method works in a given instance, the posterior [Eq. (\ref{PosteriorDist})] should show a sharp peak at the optimal $\mu^{*}$. This is equivalent to the case where the $P_{\rm P}$ of every residue is almost equal to 1 or 0. For the 3D $3\times3\times3$ lattice, however, there were several comparatively low $P_{\rm P}$ values (close to 0.5), even in the case of the highly designable conformation. Thus, our design method was not appropriate for those conformations.

The greatest advantage of our method is that it skips the exhaustive calculation of $Z(\boldsymbol \sigma)$ by assuming the prior distribution given in Eq. (13). As already stated, the form of the prior means that the lower the free energy $F_{\beta,\mu}(\boldsymbol \sigma):=-(1/\beta)\ln\Xi_{\beta,\mu}(\boldsymbol \sigma)$, the higher the prior distribution. The prior [Eq. (13)] states the hypothesis that sequences rich in P residues are, in general, more likely to be evolutionarily selected than sequences with unique stable conformations. The result of 2D $N = 16$ mentioned in Sec. III, the higher $SR$ of the non-maximally compact conformations,  is proves this hypothesis. This hypothesis is consistent with recent findings that organisms have many intrinsically disordered proteins\cite{wright1999intrinsically, uversky2000natively, dunker2001intrinsically, tompa2002intrinsically}; these proteins do not have unique native conformations and are composed of sequences rich in P residues. Recent work\cite{uversky2017intrinsically} has shown that such proteins form `droplets (that function as membraneless organelles)' and have various biologically important roles (e.g., spatiotemporal regulation of gene expression, signaling, and stress response). This indicates that organisms make good use of the physical property given by Eq. (13).

In addition to the biological validity of the prior, the fact that it enabled fast protein design without the calculation of $Z(\boldsymbol \sigma)$ is significant because it suggests that Eq. (13) is not a unique solution for protein design without exhaustive calculation of $Z(\boldsymbol \sigma)$.
As all information about the thermodynamic profile of a protein is  evolutionarily embedded solely in the sequence, it is in principle possible to search for a sequence that stabilizes a given target conformation if the code connecting $\boldsymbol \sigma$ and the thermodynamic profile is broken. The prior given by Eq. (13) may be one such code.

Finally, we discuss a simple comparison between our Bayesian design method, which introduces the chemical potential $\mu$ using the grand canonical approach, and the conventional method using the explicit solvent\cite{abeln2008disordered, ni2013interplay, abeln2014simple, bicanco2017role, bianco2019protein, bianco2020in} described in Sec. II. If we consider, for instance, the exact positions of residues that interact with water molecules or the individual value of the chemical potential for each amino acid, the $SR$ of our Bayesian method is highly expected to increase. Our method only captures the essence of the solvation of proteins using the grand canonical picture. Therefore, it presumably shows a baseline result for the above detailed implementations of the explicit solvent. The studies with direct solvent elucidate a relationship between protein aggregation and solvation\cite{abeln2008disordered, ni2013interplay, abeln2014simple, bianco2019protein, bianco2020in}, and an effect of solvation on the evolution of proteins\cite{bicanco2017role}. These are significant studies in biological physics. Especially, Bianco {\it et al.}\cite{bicanco2017role} elucidated that the segregation (the hydrophobic residues are in the core, and the polar residues are on the surface of a protein) causes instability of proteins. It gives an inspiration to our method. Therefore, a comparison study between our Bayesian design method and these studies will be an interesting future work.

\section{CONCLUSION}
The simple conclusion from these results is that it is possible to design many conformations without an exhaustive conformational search by taking the water effect into account. This approach is more successful with 2D HP models than with 3D  models; however, our method is expected to correctly design 3D target conformations given a sufficiently high designability of the target conformation. This approach is more successful for the non-maximally compact conformations.  When using 20-letter model, for the 3D $N = 3\times3\times3$, this Bayesian design method is more successful than the 2-letter (HP) model.  According to the comparison with the design method using the MTP criterion by Seno {\it et al.}\cite{seno1996optimal}, the design accuracy of the proposed Bayesian method is lower than the design accuracy achieved by Seno {\it et al.}.  By contrast, the design efficiency i.e.,  the design accuracy per the calculation time of the former significantly exceeds that of the latter.

Our Bayesian protein design method using the grand canonical approach is consistent with conventional protein design software, e.g., Rosetta. Hence, our method based on statistical mechanics may enable future studies on more realistic protein design.

Future work could consider an additional parameter reflecting the specific topological information of a target conformation. In addition, setting different numbers of water molecules to combine with each P residue would help to more closely model realistic globular proteins. As Bayesian learning is simple and flexible, such modifications could be easily implemented.

Extending our Bayesian design method for designable off-lattice proteins\cite{coluzza2011coarse, coluzza2014tranferable, hoang2004geometry, kukic2015mapping, cardelli2017role} is also an important future work. Significantly, the caterpillar model proposed by Coluzza\cite{coluzza2011coarse} is an updated model of conventional off-lattice model\cite{hoang2004geometry}. Therefore, extending our Bayesian design method proposed in this study to apply to the caterpillar model is promising.

\section*{ACKNOWLEDGMENTS}
The authors are grateful to M. Ota, Nagoya University, and Nobu C. Shirai, Mie University, for illuminating discussions.
This work was supported by KAKENHI Nos. 19H03166 (G.C.) and 19K03650 (K.T.).

\end{document}